\documentstyle[prl,epsf,aps]{revtex}
\newcommand{\be}{\begin{equation}}
\newcommand{\ee}{\end{equation}}
\newcommand{\bq}{\begin{eqnarray}}
\newcommand{\eq}{\end{eqnarray}}
\newcommand{\n}{\noindent}

\newcommand{\bc}{\begin{center}}
\newcommand{\ec}{\end{center}}

\newcommand {\al} {\alpha}

\newcommand{\g}{\gamma}
\newcommand{\gt}{\tilde{\gamma}}
\newcommand{\lt}{\tilde{\lambda}}
\newcommand{\wt}{\tilde{\omega}}

\newcommand{\G}{\Gamma}
\def \ll {\lambda}

\newcommand {\w} {\omega}

 \def\(({\left(}
 \def\)){\right)}
\def\[[{\left[}
\def\]]{\right]}

\def \la{\langle}
\def\ra{\rangle}

\begin{document}

%\begin{titlepage}

\title{{\bf Dynamical fluctuations in an exactly solvable model of
spin glasses.}}
\author{Matteo Campellone$^{*}$, Giorgio Parisi$^{**}$ and 
Paola Ranieri$^{**}$}

%\date{\today}

\address{ (*) Dipartimento di Fisica and Sezione INFN, Universit\`a 
di Padova, \\ 
 Via Marzolo 8,
 I-35131 Padova, Italy \\
 (**) Universit\`a di Roma ``La Sapienza''\\
 Piazzale A. Moro 2, 00185 Rome (Italy)\\
 e-mail:{\it  campellone@roma1.infn.it, 
giorgio.parisi@roma1.infn.it, paola.ranieri@roma1.infn.it}}

\date{\today}
\maketitle
\vspace{2truecm}

\begin{abstract}
\n
In this work we calculate the dynamical fluctuations at O(1/N) in the low 
temperature phase of the $p=2$ spherical spin glass model. We study the 
large-times asymptotic regimes and we find, in a 
short time-differences regime, a fluctuation dissipation relation 
for the four-point correlation functions. This relation can be 
extended to the out
of equilibrium regimes introducing a function $X_{t}$ which, for large 
time $t$, as $t^{-1/2}$ as in the case of the two-point functions. 
\end{abstract}

%\end{center}
\vspace{1.cm}
\hspace{.1in} PACS Numbers 05-7510N

%\end{titlepage}

\section{Introduction}

The mean field Langevin dynamics for spin glasses has been quite 
extensively studied in recent years \cite{s.z.2}, \cite {s.}, 
\cite{p-spin}, \cite{sk}, \cite {rm},\cite{rm2}, \cite{cu.de.}. 
The main result has been that for low enough temperatures there 
is an off-equilibrium regime where the dynamics of the 
system depends on its whole history up to the beginning of its 
observation and often this feature is accompanied by a loss of the validity 
of the fluctuation dissipation theorem (FDT). 
Though it is in general very difficult to calculate the explicit time
dependence of the correlation functions, except for some simple models,  
\cite{cu.de.}, their behaviour in all the different asymptotic 
regimes has been well understood \cite{p-spin} \cite{sk}.

The main question we are left with is how to extend the mean field picture 
for finite dimensional systems. From the analytical point of view this amounts to 
take into account the corrections to the mean field limit.

In all generality one can determine closed equations for the 
two-point correlation functions as saddle points of an appropriate 
functional of a two time dependent field 
$Q_i^{\alpha\beta}(t_1,t_2)$, \cite{s.z.2}. In this formalism, to 
consider the dynamical fluctuations around this mean field limit,  
one has to solve the equations of motion for the propagators, which 
are some four-times correlation functions and are related to the 
dynamical spin glass susceptibility.
 
At the dynamical critical temperature, where dynamical scaling is 
supposed to hold and the off-equilibrium features are not relevant, these 
equations have been solved for some models \cite{pro}. 

Unfortunately 
it is not clear how to approach the equations of motion below $T_{c}$ 
unless one knows which are the large-time asymptotic regimes for 
these four-time functions at low temperatures.

Furthermore one would like to understand if, and in which time regimes, 
a kind of fluctuation-dissipation relation can be written for these 
four-time dependent quantities.

These questions can by answered in the case of a simple spin-glass 
model which can be solved explicitly.
It describes soft spins interacting through quenched random long
range couplings and forced to satisfy a global spherical constraint.
Statically,  this so called {\em spherical $p=2$ spin glass} 
does not present the peculiar features typical
of other spin glasses and reveals itself to be a sort of disguised
ferromagnet. Below a critical temperature $T_c$ the system freezes in 
one of two possible states spontaneously breaking the symmetry under the
parity operator that connects them. 

Nevertheless, for temperatures below $T_c$ and if the dynamics of
the system begins in a random 
initial configuration, the model presents a 
non trivial out of equilibrium dynamical regime exhibiting the so 
called aging phenomenon and a violation of the fluctuation-dissipation 
theorem.

The simplicity of this model allowed for an explicit calculation of 
the two-point
correlation and response functions \cite{cu.de.}.
In this work we shall calculate the explicit form of the four-point
correlation functions.
Hopefully, this simple case will provide us some hints on how to approach more 
complicated models which cannot be solved without the use of the 
functional methods and therefore without an ansatz on the asymptotic 
behaviour of the solution. 

\section{The Statics}

The spin glass spherical model is described by the Hamiltonian
\be
H = -{1\over 2}\sum_{i\neq j} J_{ij} s_i s_j - \sum_i h_i s_i ,
\label{ham}
\ee
where $h_i(t)$ is an 
external magnetic field,
and the spin variables ${s_i}$ are forced to satisfy the spherical constraint

\be
\sum_i^N s_i ^2 = N.
\ee

The couplings $J_{ij}$ are symmetric quenched random variables extracted 
from a Gaussian probability distribution with $\overline{J_{ij}}=0$ and 
$\overline{J_{ij}^2}=1/N$.
The mean field solution of the statics is rather simple for this
model \cite{ko.to.}.
Note that, using conventional notations, we have indicated by $\la . 
\ra$ the averages over the Boltzmann distribution, and by 
$\overline{.}$ the averages over the different realizations of the 
quenched disorder $J_{ij}$.

The partition function is
\be
Z = \int_{-\infty}^{\infty} \frac{dz}{2\pi i} \prod_{i}^{N} ds_i
\exp{\left(\frac{\beta}{2} \sum_{i\neq j} J_{ij} s_i s_j - z \sum_i s_i^{2}
\right)},
\ee

where we have set $h_i=0$ for all $i$ and $z$ is a Lagrange multiplier 
introduced to enforce the constraint.
For $N \rightarrow \infty$ the distribution of the 
eigenvalues of the matrix $J_{ij}$ follows the 
Wigner semi-circle law \cite{Me}:

\be
 \rho(\ll) = {1\over 2 \pi} \sqrt{4 - \ll^2} \;,  \;\;\;\;\;\;\;
\ll \in [-2,2].
\label{rho}
\ee
In this limit one can formally write the saddle point equations on $Z$ 
which reproduce a softer version of the constraint

\be
1 = \frac{1}{N} \sum_{\al}^{N} \la s_{\al}^2 \ra  = \int_{-2}^{2} d 
\ll
\rho(\ll) \frac{1}{2z - \beta \ll}.
\label{cmf}
\ee

In equation (\ref{cmf}) $s_{\al}=\sum_{i=1}^{N} \phi_i^{\alpha} s_i$ 
is the projection of the spin variables on the $\alpha$-th eigenvector of the 
matrix of the couplings $J_{ij}$.

It can be seen that below a critical temperature $T_{c}=1$
the second of the equalities in (\ref{cmf}) does not hold because a 
spontaneous magnetization arises along the eigenvector with eigenvalue 
2 (or $-2$). For $T<T_c$ the value of the Lagrangian multiplier remains
fixed at the branch point $z=\beta$ and a magnetization 
$\la s_2\ra\propto N^{1/2}$ appears.

The spin glass susceptibility is defined as follows:
\be
\chi_{SG}  = \frac{1}{N}\sum_{ij} \overline{\(( \la s_{i} s_{j} \ra
-\la s_{i} \ra \la s_{j} \ra \))^2 } = \frac{1}{N} \sum_{\al}
\frac{1}{(2 z - \beta \ll^{\al})^2},
\ee

In the large-$N$ limit $\chi_{SG}$ diverges as $1/(T-T_c)$ 
for $T\rightarrow T_c^{+}$ and it remains infinite in the whole frozen 
phase. This model is critical for all temperatures below $T_{c}$. 

The computation of the average free energy can also be done using
the replica method, \cite{ci.dp.}, where a replica symmetric ansatz solve 
the model exactly.  One can reproduce the results for $\chi_{SG}$
using the following identity:
\be
\chi_{SG} = \lim_{n \rightarrow 0} \[[ \langle \delta q_{\al \beta}^2
\rangle -2 \langle \delta q_{\al \beta} \delta q_{\al \g} \rangle +
\langle \delta q_{\al \beta} \delta q_{\g \delta } \rangle \]],
\label{chirep}
\ee
where the $\delta q_{\al \beta}$ are the fluctuations around the
saddle point of the replicated partition function.
It can be seen that each of the terms of equation (\ref{chirep}) diverges
below $T_{c}$.

\section{The Mean Field Dynamics}

We shall mainly deal with the dynamical behaviour of the model. 
A Langevin dynamics for the Hamiltonian (\ref{ham}) joined with the
spherical constraint gives the following equation of motion
\be
\frac {\partial s_i(t)}{ \partial t} = \sum_{j} J_{ij} s_j(t) - 
z(t) s_i(t) +h_i (t) + \xi_i(t)
\label{lang}
\ee
where $z(t)$ is a time-dependent Lagrange multiplier,
 and $\xi_i(t)$ is a Gaussian noise with zero mean 
and variance $\la\xi_i(t)\xi_j(t')\ra=2T\delta_{ij}\delta(t-t')$.
In this model the solution of (\ref{lang}) for the component $s_{\alpha}$ 
can be explicitly written
\be
s_\al (t) = s_{\al(t_{0}) } e^{\al (t-t_{0}) -\int_{t_{0}}^{t}
z(\tau)
d\tau    } + \int_{t_{0} }^{t}e^{\al(t-t'') -\int_{t_{0}}^{t} z(\tau')
d\tau' } (h_{\al}(t'')+\xi_{\al}(t''))dt'' ,
\label{csol}
\ee 
where again $\al$ labels the eigenvectors of $J_{ij}$, and the initial 
time is $t=0$.
We introduce the correlation and response function which read:

\bq
C(t_{1},t_{2}) & = & \frac{1}{N} \sum_{i}^{N}\la s_{i}(t_{1})
s_{i}(t_{2}) \ra, \\
G(t_{1},t_{2})& = & \frac{1}{N} \sum_{i}^{N}
 {\partial\la s_i(t_{1} )\ra \over \partial h_i(t_{2} )}= 
\frac{1}{N} \sum_{i}^{N}
 {\partial\la s_i(t_{1} )\ra \over \partial \xi_i(t_{2} )}=
\frac{1}{N} \sum_{i}^{N} \frac{1}{2 T} \la s_i(t_{1} )\xi_i(t_{2} )\ra ,
\label{corrs}
\eq
where the last equality is valid because the noise has a Gaussian
distribution probability. 
In the $N\rightarrow\infty$ limit and in absence of 
the external field the above functions have the form 
\cite{cu.de.}

\bq
C(t_{1},t_{2}) & = & \frac{1}{\sqrt{ \G(t_1)\G(t_3)}} \[[ \frac{I_{1}
 [2 (t_{1} + t_{2})]}
{(t_{1} + t_{2})} + 2 T \int_{0}^{t_{2} }\! \!
\!dt' \,
\G(t')  \frac{I_{1} (t_{1}+t_{2}-2t')}{(t_{1}+t_{2}  - 2t')}
\]] \label{mfd1}\\
G(t_{1},t_{2})& = & \sqrt{\frac{ \G(t_1)}{\G(t_2)}}
\frac{I_{1} (t_{1}-t_{2})}{(t_{1}-t_{2})},
\label{mfd2}
\eq
where $I_1[x]$ is the modified  Bessel function and the function
\be
\G(t) \doteq e^{2 \int_{t_{0}}^{t} z(\tau) d\tau}
\ee
is fixed by implementing the spherical condition
$C(t,t)=1$ in equation (\ref{mfd1}).

The simplicity of this model allows one to obtain the mean field 
solutions  (\ref{mfd1}) (\ref{mfd2}) directly, simply averaging over the 
distribution of the eigenvalues of the matrix $J_{ij}$.

For other models of spin glasses we do not manage to obtain the 
explicit form of $C(t_{1},t_{2})$ and  $G(t_{1},t_{2})$.

A quite general procedure, introduced in  \cite{s.z.2}, allows to obtain 
closed equations for $C(t_{1},t_{2})$ and $G(t_{1},t_{2})$ as the saddle 
point solutions of a dynamical generating functional. 
In particular, in the case the off-equilibrium dynamics of the $p$-spin model,
with $p>2$, and the SK model, we are not able to solve
these equations and therefore to determine 
the explicit time dependence of the correlation and response 
functions but we can only predict the structure of their asymptotic behaviour. 
This model that provides us with an explicit solution,  
allows us to control the assumptions and the ansatz we used to obtain 
analytical results for the dynamics of more complicated systems. 

For the correlation functions $C(t_1,t_2)$ and $G(t_1,t_2)$ it has
been shown that there are basically two asymptotic time scales and that
they can be distinguished by the variable $\ll = t_2/t_1$ 
\cite{cu.de.}.

$\bullet$ For $\ll \simeq 1$ the system is in an equilibrium regime, 
in which the functions depend only on the difference of the 
two arguments (time translational invariance) and the FDT relation

\be
G(t_1-t_2) = \frac{1}{T} {\partial C(t_1-t_2) \over \partial t_2}
\theta(t_1-t_2), 
\ee

holds at all temperatures. 

$\bullet$ For $\ll \simeq O(1) < 1$ and for $T < T_c$, the system is
in the so called {\em aging} regime: the correlation functions depend on both 
time variables
(in this case through $\ll$) and not only on time differences. 
Moreover the FDT relation is not valid although it can be generalized
introducing a function $X_{t_1}(C)$ \cite{p-spin} \cite{sk} by the relation

\be
G(t_1,t_2) = \frac{X_{t_1}(C)}{T} {\partial C(t_1,t_2) \over \partial t_2}
\theta(t_1-t_2). 
\ee
For large times one has that $X_{t_1}(C)\rightarrow X(C)$. The function 
$X(C)$
generally characterizes the  type of aging dynamics of the model. 
In this model the slowness of the dynamics is due to the flatness of the 
energy landscape and is therefore qualitatively similar to ordinary 
domain coarsening. 
This kind of dynamics is in general associated with a ${X(C)}=0$ which is 
indeed the case for this model in which it has been found that, in the aging 
regime, $X_{t_1}(C)$ scales as $t_1^{-1/2}$ for large $t_1$, \cite{cu.de.}.
We shall see that this will be the case also for the four-times 
functions.

\section{Dynamical Fluctuations} 
 
The problem of the dynamical fluctuations around the mean field 
solution for spin glass models, for $T<T_c$ has not been faced yet.
We shall therefore study the four-point functions in this model in
which it is possible to determine the explicit temporal behaviour 
of these functions and we will describe the different large-time 
asymptotic regimes.  
 
Let us now introduce the four-time correlation functions 

\bq
\g(t_1,t_2,t_3,t_4) &\doteq&\frac{1}{N^2}\sum_{i,j}\Big( 
\overline{\la s_{i} (t_1)
s_{i} (t_2)  \xi_{j} (t_3) \xi_{j} (t_4) \ra }- \overline{\la s_{i} 
(t_1) s_{i} (t_2) \ra}\,\overline{
 \la \xi_{j} (t_3) \xi_{j} (t_4) \ra }\Big)\label{gamma} \\
\ll(t_1,t_2,t_3,t_4) &\doteq& \frac{1}{N^2}\sum_{i,j}\Big(
\overline{\la s_{i} (t_1) 
s_{i} (t_2) s_{j} (t_3) \xi_{j} (t_4) \ra }- \overline{ \la s_{i} (t_1)
s_{i} (t_2) \ra}\,\overline{ \la s_{j} (t_3) \xi_{j} (t_4) \ra }\Big)
\label{lambda} \\
\w(t_1,t_2,t_3,t_4) &\doteq& \frac{1}{N^2}\sum_{i,j}\Big(\overline{ \la 
(s_{i} (t_1)- s_{i} (t_2))^2
(s_{j} (t_3)- s_{j} (t_4))^2 \ra}-  \nonumber\\
&&\hspace{2cm} \overline{\la (s_{i} (t_1)-
s_{i} (t_2))^2 \ra}\,\overline{
 \la (s_{j} (t_3)- s_{j} (t_4))^2 \ra }\Big).
\label{omega}
\eq

The functions
$\g, \ll, \w$ are related to the fluctuations around the saddle 
point of $Q_i^{\alpha\beta}(t_1,t_2)$, \cite{s.z.2} \cite{s.} that gives, 
as mean field solutions, the two-point correlations and response 
functions. 
Note that $\g(t_1,t_2,t_3,t_4)$ is related to the dynamical
$\chi_{SG} $:
\be
\chi_{SG}=\frac{1}{N}\sum_{ij}\overline{\frac{\partial \la s_i(t_1)\ra }
{\partial h_j(t_3)}
\frac{\partial \la s_i(t_2)\ra }{\partial h_j(t_4)}}\;,
\ee
while $\w(t_1,t_2,t_3,t_4)$ is the dynamical four-points 
correlation function defined in such way to stay finite in the asymptotic
limit of $t_1\rightarrow \infty$ with $t_1\sim t_2\sim t_3\sim t_4$, i.e.
on the equilibrium time scale. 

In this regime one expects to verify some FDT-like relations
\bq
\frac{\partial}{\partial t_4}  \w(t_1,t_2,t_3,t_4) & = & 
2T \Big(2 \ll(t_1,t_2,t_3,t_4)-\ll(t_1,t_1,t_3,t_4)
-\ll(t_2,t_2,t_3,t_4)\nonumber\\
&&-2\ll(t_1,t_2,t_4,t_4)+
\ll(t_1,t_1,t_4,t_4)+\ll(t_2,t_2,t_4,t_4) \Big), \label{fdt21}\\  
\frac{\partial}{\partial t_3} \ll(t_1,t_2,t_3,t_4) & =& T 
\g(t_1,t_2,t_3,t_4).
\label{fdt22}
\eq

Let us remark that in equation (\ref{gamma}) (\ref{lambda}) 
(\ref{omega}) we have by hand subtracted the part of the term
that is of order $O(1)$, and the quantities defined are all of 
order $O(1/N)$. 

In the functional formalism the above functions are the propagators 
of the fields $Q^{\alpha\beta}(t_1,t_2)$ introduced in \cite{s.z.2}
evaluated at zero momentum.  
This simple model will provide us with the first explicit calculation 
of the dynamical propagators for a spin glass in the low temperature phase. 

Let us now set the time order $t_4<t_3<t_2<t_1$.
Using the fact that the model is quadratic we can write   
\bq
\g(t_1,t_2,t_3,t_4)&=& \gt(t_1,t_2,t_3,t_4)+\gt(t_1,t_2,t_4,t_3)\\
\ll(t_1,t_2,t_3,t_4)&=& \lt(t_1,t_2,t_3,t_4)+\lt(t_2,t_1,t_3,t_4)\\ 
\w(t_1,t_2,t_3,t_4)&=& 4 \wt(t_1,t_2,t_3,t_4)+4 \wt(t_1,t_2,t_4,t_3)-4 
\wt(t_1,t_1,t_3,t_4)-\nonumber\\
&& \hspace{-1cm} 4 \wt(t_2,t_2,t_3,t_4)
-4 \wt(t_1,t_2,t_3,t_3)-4 \wt(t_1,t_2,t_4,t_4)+\nonumber\\
&&\hspace{-3cm} 2 \wt(t_1,t_1,t_3,t_3)+ 2 \wt(t_2,t_2,t_3,t_3)+
2\wt(t_1,t_1,t_4,t_4)+2 \wt(t_2,t_2,t_4,t_4)
\eq
where we have defined
\bq
\gt(t_1,t_2,t_3,t_4) &\doteq& \overline{\la s_{i} (t_1)
\xi_{j} (t_3)\ra \la  s_{i} (t_2) \xi_{j} (t_4) \ra } \\
\lt(t_1,t_2,t_3,t_4) &\doteq&  \overline{\la s_{i} (t_1)
s_{j} (t_3)\ra \la  s_{i} (t_2) \xi_{j} (t_4) \ra } \\
\wt(t_1,t_2,t_3,t_4) &\doteq&  \overline{\la s_{i} (t_1)
s_{j} (t_3)\ra \la  s_{i} (t_2) s_{j} (t_4) \ra }.
\eq

One gets

\bq
\gt(t_1,t_2,t_3,t_4) &=&\frac{1}{N} \theta(t_1 - t_3)
\theta(t_2-t_4)
 \sqrt{\frac{\G(
t_3)\G(t_4)}{\G(t_1)\G(t_2)}} \frac{I_1[2 (t_{1} + t_{2} 
-t_{3}-t_{4})]}
{(t_{1} + t_{2} -t_{3}-t_{4})}  \label{fz4pti1}  \\
\lt(t_1,t_2,t_3,t_4) &=&\frac{1}{N}
\theta(t_2-t_4) \frac{ \sqrt{\G(t_4)} }{\sqrt{\G(
t_1)\G(t_2)\G(t_3)}}\left[\frac{I_1[2 (t_{1} +
t_{2}+t_{3}-t_{4})]}{(t_{1} + t_{2} +t_{3}-t_{4})}\right.\nonumber\\
&&\left.+ 2 T \int_{0}^{t_{3} } dt' \G(t')\frac{I_1[2 (t_{1} +
t_{2}+t_{3}-t_{4}- 2 t')]}{(t_{1} + t_{2} +t_{3}-t_{4} - 2 t')}
\right]\label{fz4pti2}\\
\wt(t_1,t_2,t_3,t_4) & = & \frac{1}{N}
\frac{1}{\sqrt{\G(
t_1)\G(t_2)\G(t_3)\G(t_4)}}\times \nonumber\\
&&\hspace{1cm}\left[ 2T \int_{0}^{t_{3} } \! \!\!dt'
 \G(t') \frac{I_1[2 (t_{1}+t_{2}+t_{3}+t_{4}-2t')]}
{(t_{1}+t_{2}+t_{3}+t_{4}-2t')}+\right.\nonumber\\
&&\hspace{.5cm}2T \int_{0}^{t_{4} } \! \!\!dt'
 \G(t') \frac{I_1[2 (t_{1}+t_{2}+t_{3}+t_{4}-2t')]}
{(t_{1}+t_{2}+t_{3}+t_{4}-2t')}+\nonumber\\
&&\hspace{-2cm} \left. 4 T^2
\int_{0}^{t_{3} }\! \!
\!dt'\!\int_{0}^{t_{4} } \! \!\!dt''
 \G(t')  \G(t'')\frac{I_1[2 (t_{1}+t_{2}+t_{3}+t_{4}-2t'-2t'')]}
{(t_{1}+t_{2}+t_{3}+t_{4}-2t'-2t'')}\right]. 
\label{fz4pti3}
\eq

We shall study the fluctuations around the mean field solutions and 
in particular we shall be interested in understanding if equations 
(\ref{fdt21}) and (\ref{fdt22}) are verified. 

In the large-time limit form the functions (\ref{fz4pti1})(\ref{fz4pti2})
(\ref{fz4pti3}) read
\bq
\gt(t_1,t_2,t_3,t_4) & = &  \[[\frac{t_1 t_2}{t_3 
t_4}\]]^{\frac{3}{4}}
e^{-2 [t_{1} + t_{2} -t_{3}-t_{4}]} \frac{I_1[2 (t_{1} + t_{2} 
-t_{3}-t_{4})]}
{(t_{1} + t_{2} -t_{3}-t_{4})}    \\
\lt(t_1,t_2,t_3,t_4)\!& = &\!\! \[[\frac{4(t_1 t_2 
t_3)}{t_4 (t_1 +t_2 +t_3 -t_4)^2}\]]^{\frac{3}{4}} \nonumber\\
&& \[[1 - T  \int_{0}^{2 (t_1 +t_2 -t_3 -t_4) }  dt' 
\frac{e^{-t'}I_1[ t']}{t' (1-\frac{t'}{2(t_1 +t_2 +t_3 
-t_4)})^{(\frac{3}{2})}
}\]] \\
\wt(t_1,t_2,t_3,t_4)\!& =&  q_{EA}^2 \sqrt{4 \pi} 
\[[ \frac{16( t_1 t_2 t_3 t_4)}{(t_1+ t_2+ t_3 +t_4)^2} 
\]]^{\frac{3}{4}}\big[ 2(1-q_{EA}^2)- 2T+2T\times\nonumber\\
&&\hspace{-3.5cm}\left. 
 \int_{0}^{t_3}\!\!\!\! dt'' \frac{\G(t'')e^{-4 t''}}{(1-\frac{2 t''}
{t1+t2+t3+t4})^{3/2}} 
\[[1 -q_{EA}^2 -T \int_{0}^{2 (t_1 +t_2 +t_3 -t_4 -2 t'' ) }\!\!\!\!
 dt' \frac{e^{-t'}I_1[t']}{[t' (1-\frac{t'}{2(t_1 +t_3 +t_2 +t_4 - 
2t'')})^{3/2}]} \]]\]]\label{omega1}
\eq
Note that the function $\wt$ is divergent as $t_1^{3/2}$ in all the time 
regimes. This is due to the fact that $\wt(t_1,t_2,t_3,t_4)$ diverges
in the static limit.

However, in the definition (\ref{omega}) of the $\w(t_1,t_2,t_3,t_4)$ 
we have subtract this overall behaviour on the equilibrium time scale,  
and in this regime we can consider only the dynamics on the time 
differences.  

The structure of the possible time regimes is in principle similar to 
that for the two point correlation functions: the times are either far 
or close to each other.
 
In a generic four-time function we can have some times which are 
close and some which are far away from each other and this 
complicates the separation in time sectors of the time space because 
the function may be in local equilibrium with respect to some
times and aging with respect to others.
We will now study the functions in all the possible times regimes 
obtainable by four ordered times. We will calculate their asymptotic 
behaviour for each regime. The result of our analysis is
 that relations (\ref{fdt21}) and (\ref{fdt22}) hold if and only if all the 
four times are close to each other. That 
means if there are no aging time scales. 
Otherwise the dependence upon the aging times 
({\em i.e.} they are far from each other) will dominate the functions.  
 
The results that we obtain are summarized in the table below
 where with $\sum\ll$
we indicated in a compact form the r.h.s. of the equation 
(\ref{fdt21}).On the first column we indicated the time regime we were 
considering, using curly brackets to group times whose difference 
remains finite in the limit of $t_{1} \rightarrow \infty$.

\medskip
\begin{equation}
\begin{array}{|c|c|c|c|c|c|c|} \hline       
\par
Time Regime &\w& \frac{\partial\w}{t_4}&
\sum\ll&\ll& 
\frac{\partial\ll}{t_3}&\g\\ \hline   
(t_1 t_2 t_3 t_4)& t_1^{-5/2} &T \sum\ll& O(1) &
O(1)&T \g& O(1)  \\ \hline  
(t_1 t_2)(t_3 t_4)& t_1^{-5/2} &t_1^{-5/2} &t_1^{-3} &
O(1) &t_1^{-1} & t_1^{-3/2}\\ \hline 
t_1 t_2 (t_3 t_4)& t_1^{-1/2} &t_1^{-1/2} &t_1^{-1} &
O(1) &t_1^{-1} & t_1^{-3/2}\\ \hline 
t_1 (t_2 t_3) t_4& t_1^{3/2} &t_1^{1/2} &O(1) &
O(1) &t_1^{-1} & t_1^{-3/2}\\ \hline 
(t_1 t_2) t_3 t_4 & t_1^{-1/2} &t_1^{-3/2} &t_1^{-2} &
O(1) &t_1^{-1} & t_1^{-3/2}\\ \hline 
(t_1 t_2 t_3) t_4& t_1^{-1/2} &t_1^{-3/2} &t_1^{-2} &
O(1) &t_1^{-1} & t_1^{-3/2}\\ \hline 
t_1 (t_2 t_3 t_4)& t_1^{-1/2} &t_1^{-1/2} &t_1^{-1} &
O(1) &t_1^{-1} & t_1^{-3/2}\\ \hline 
t_1 t_2 t_3 t_4& t_1^{3/2} &t_1^{1/2} &O(1) &
O(1) &t_1^{-1} & t_1^{-3/2}\\ \hline 
\end{array}
\nonumber
\end{equation}
\medskip

To see if equations (\ref{fdt21}) and (\ref{fdt22}) hold one has to compare 
the third column with the fourth and the sixth column with the seventh.   
One sees that the FDT relations for the fluctuations hold only in 
the first regime, where all the times are at a finite distance from 
each other. In the other regimes they still can be generalized 
introducing an $X_{t_{1}}(C)  \propto t_{1}^{-1/2}$ as in the case of the 
two-times functions.
Let us also remark that $\w(t_{1},t_{2},t_{3},t_{4})$ diverges every 
time  $t_1-t_2 $ and $t_3-t_4$ are order $O(t_{1})$. In this cases, 
the function $\w(t_{1},t_{2},t_{3},t_{4})$, (\ref{omega}),
has not a well defined limit.
Moreover, in the FDT regime the functions  $\g(t_{1},t_{2},t_{3},t_{4})$,   
$\ll(t_{1},t_{2},t_{3},t_{4})$ and $\w(t_{1},t_{2},t_{3},t_{4})$ 
are asymptotically
dependent on the times only trough $\tau=t_{1}+t_{2}-t_{3}-t_{4}$, 
while in the aging regimes an explicit dependence from the times 
that are at distance $O(t_1)$ remains. 
Finally  we would like to point out that $\tau$ is a characteristic 
time scale that determines whether we are in an FDT regime 
($\tau\sim O(1)$) or not ($\tau\sim O(t_1)$), 
assuming the same role as $t-t^{\prime}$ in the case of the two times 
functions. 

\section{Conclusions}  

In this work we calculated explicitly the four-times functions
for the spherical $2$-spin glass model in the low temperature phase.
We calculated the asymptotic behaviour of these functions in the possible 
time regimes that can be selected with four ordered times. 
We found out that there is a time regime in which one can relate the 
functions through a relation similar to the fluctuation-dissipation 
theorem for the equilibrium dynamics of the two-times functions.
In this work we provide the first explicit expression for the dynamical 
four-point functions for a spin glass model in the cold phase.
It is our intention  to use the results obtained here 
to analyze less trivial spin glass models describing short range 
interactions of $p>2$ spins.  
In these models there is lot to be understood about the dynamical 
propagators and about their static limit \cite{mapagi}. Unfortunately the
equations for the propagators are integro-differential equations that 
cannot be solved without the use of an ansatz on their asymptotic form.
We think that the knowledge of the dynamical structure of the 
propagators for the spherical $p=2$ spin glass model can help us in view
of these further developments.

\end{document}